\documentclass[11pt]{article}

\usepackage{amsmath, graphics, graphicx, latexsym, amssymb, amsthm, amsfonts}
\usepackage[mathscr]{eucal}
\usepackage{epsfig}
\usepackage[english]{babel}
\usepackage[square]{natbib}
\usepackage{setspace, enumerate}
\usepackage[pdfauthor={AG}]{hyperref}
\newtheorem{theo}{{\bf{Theorem}}}[section]
\newtheorem{prop}[theo]{{\bf Proposition}}
\newtheorem{lem}[theo]{{\bf Lemma}}

\newtheorem{rem}{{\bf Remark}}[section]

\newtheorem{remark}[rem]{Remark}
\newtheorem{defi}{{\bf Definition}}[section]
\newtheorem{defn1}[defi]{Definition}

\renewcommand{\proof}{\noindent{\bf Proof.\ }}
\newcommand{\no}{\nonumber}
\newcommand{\noi}{\noindent}
\newcommand{\txt}{\textrm}

\newcommand{\ds}{\displaystyle}

\newcommand{\si}{\sigma}

\newcommand{\calF}{\mathcal{F}}

\newcommand{\sig}{\sigma}
\newcommand{\tab}{\hspace*{0.3in}}

\newcommand{\vf}{\varphi}

\catcode`\@=11
\catcode`\@=12

\begin{document}

\title{The optimal hedging in a semi-Markov modulated market}
\author{Anindya Goswami\thanks{IISER, Pune 411008, India; email: anindya@iiserpune.ac.in}\qquad\qquad Jeeten Patel \thanks{IISER, Pune 411008 India; email: jeeten@iiserpune.ac.in}\qquad\qquad Poorva Shevgaonkar \thanks{IIT Kharagpur, India; email: poorvashevgaonkar@gmail.com}\\ }


\maketitle

{\bf Abstract:}
This paper includes an original self contained proof of well-posedness of an initial-boundary value problem involving a non-local parabolic PDE which naturally arises in the study of derivative pricing in a generalized market model. We call this market model a semi-Markov modulated market. Although a wellposedness result of that problem is available in the literature, but this recent paper has a different proof. Here the existence of solution is established without invoking mild solution technique. We study the well-posedness of the initial-boundary value problem via a Volterra integral equation of second kind. The method of conditioning on stopping times was used only for showing uniqueness. Furthermore, in the present study we find an integral representation of the PDE problem which enables us to find a robust numerical scheme to compute derivative of the solution. This study paves for addressing many other interesting problems involving this new set of PDEs. Some derivations of external cash flow corresponding to an optimal strategy are presented. These quantities are extremely important when dealing with an incomplete market. Apart from these, the risk measures for discrete trading are formulated which may be of interest to the practitioners.


{\bf Classification No.:} 60K15, 91B30, 91G20, 91G60.
\section{Introduction}
In the literature of derivative pricing, consideration of an incomplete market as underlying, is quite common. In such a market there may be no self financing hedging strategy which can replicate a given legitimate claim at the maturity. Hence pricing problem is rather involved. Nevertheless for an incomplete market, there are several different approaches to formulate a price. Here we focus on the local risk minimization approach as given in \cite{FSO}; \cite{S1}; \cite{S}; \cite{S3} and \cite{S2}. In this approach, to hedge a claim, theoretically, one adopts a particular dynamic strategy which replicates the claim at the maturity by allowing additional cash flow while performing a continuous trading. This particular strategy is the one which minimizes a certain measure of the accumulated cash flow, more specifically it minimizes a functional known as \emph{quadratic residual risk} (QRR) under a certain set of constraints. This specific minimizing strategy is known as the \emph{optimal hedging}. Existence of such strategy is derived in the literature mention above. It is also shown in \cite{FS} that the existence of an optimal hedging is equivalent to that of F\"{o}llmer Schweizer decomposition of the relevant discounted claim. Most interestingly the price function in the above sense often satisfies a well posed Cauchy problem provided the underlying market is a suitable generalization of \emph{geometric Brownian motion} GBM. Therefore, for some particular market models one can actually solve associated differential equation to obtain the price and optimal hedging of a suitable claim. And thus, one can compute the minimized QRR.

To retain mathematical tractability we consider a reasonably general class of market models which also includes GBM and Markov modulated GBM model as special cases. A regime switching market model is one where the market parameters are assumed to vary with time as a stochastic process with finite states. Nevertheless, the dependence of underlying assets on these parameters are similar to that of GBM \cite{BAS}; \cite{BE}; \cite{DES}; \cite{DKR}; \cite{GZ}; \cite{MR} and \cite{RR}. We consider a regime switching market model where the parameters follow a semi-Markov process. We call this model a semi-Markov modulated GBM model. There are some statistical results in the literature (see \cite{JH} and the references therein for more details) which emphasize the advantage of use of semi-Markov switching models over simple homogeneous Markov switching models. In fact memoryless property of Markov processes is rather restricted whereas, semi-Markov processes provide interesting simple and flexible alternative. For example it is mainly useful to deal with the impact of a changing environment (i.e. the business cycle), which exhibits duration dependence. This motivates us to consider this generalization.

Option pricing in a semi-Markov modulated market using F\"{o}llmer Schweizer decomposition is studied in \cite{AGMKG}. There it is shown that the price function satisfies a non local system of parabolic PDE. In this paper we show that the same price function also satisfies a Volterra integral equation of second kind. Furthermore we show that the PDE in \cite{AGMKG} is equivalent to the Volterra equation of this paper. In the Subsection 3.2 we present a rigorous proof of existence and uniqueness of the solution of both of the equations and their equivalence. In many papers, dealing with regime switching markets, a special case of this PDE arises \cite{BAS}; \cite{DES}; \cite{DKR} and \cite{MR}. Owing to the simplicity of the special case, generally authors refer to some standard results in the theory of parabolic PDE for existence and uniqueness issues. But in its general form which arises in this paper, no such ready reference is available. So, we produce a self contained proof using Banach fixed point theorem.

Apart from a purely mathematical interest, the results presented in the Subsection 3.2 also help to compute price and hedging functions numerically. A robust numerical scheme for hedging is derived from the integral equation. This in turn indues us to study the optimal hedging from the perspective of a discrete trader. Although in reality only the discrete trading takes place, but for the sake of mathematical tractability, continuous trading is assumed in most of the mathematical models. Owing to the mathematical intractability of discrete trading, we consider a real trader who performs discrete trading by closely following price and hedging suggestions from the theory of continuous trading. Needless to mention that such practice must lead to nonreplicability of the claim and a greater extra cash flow than that obtained in the theory of continuous trading. We find it interesting to formulate a risk measure which takes care of both, and can be computed for any specific market. We call that as practitioners measure (PM). PM is not a particular risk functional but a particular measurement of given risk functional by taking care of occurrence time of discrete trading. Our formula for PM, involves the optimal hedging function. Hence it is essential to have a robust numerical method of computing optimal hedging.

In this paper, we also compute QRR of optimal hedge of a European call option in a semi-Markov modulated GBM market. We notice that although consideration of QRR as a measure of cash flow makes the problem mathematically tractable, but it overestimates the actual risk. The actual risk is generally regarded as the NPV(net present value) of borrowing external money. To study the difference, we introduce a functional which measures only the cash inflow. We compute this quantity corresponding to the optimal hedge and compare that with QRR.

The rest of this paper is arranged in the following manner. Following \cite{FS} we present a brief description of locally risk minimizing hedging in a general incomplete market in Section 2. The description of the market model, under consideration, is presented in Section 3. This section also contains the derivation of risk minimizing price and hedging of European call option. In section 4, formulae of different risk measures are obtained for this specific market model. Finally Section 5 deals with computational aspects of the theoretical formulae.

\section{Preliminaries}
Let a market consist of two assets $\{S_t\}_{t\geq0}$ and $\{B_t\}_{t\geq0}$ where $S_t$ and $B_t$ are continuous semi-martingales and $B_t$ is of finite variation.
An \emph{admissible strategy} is a dynamic allocation to these assets and is defined as a predictable process $\pi=\{\pi_t=(\xi_t,\varepsilon_t), 0\leq t\leq T\}$ which satisfies conditions, given in $(A1)$ below.
The components $\xi_t$ and $\varepsilon_t$ denote the amounts invested in $S_t$ and $B_t$ respectively at time $t$.
The value of the portfolio at time t is given by
\begin{eqnarray}\label{1a}
V_t = \xi_tS_t + \varepsilon_tB_t.
\end{eqnarray}
\noi Here we assume
\begin{itemize}
\item[(A1)](i) $\xi_t$ is square integrable w.r.t $S_t$,\\
(ii) $E(\varepsilon^2_t)<\infty$,\\
(iii) $\exists a>0$ s.t. $P(V_t\geq-a,t\in[0,T])=1$.
\end{itemize}

Let $C_t$ be the accumulated additional cash flow due o a strategy $\pi$ at time $t$. Then $V_t$ can
also be written as sum of two quantities, one is the return of the
investment at an earlier instant $t-\Delta$ and the other one is the instantaneous cash flow $(\Delta C_t)$.
\begin{eqnarray}\label{eq.1.1}
ie. \quad V_t &=& \xi_{t-\Delta}S_t + \varepsilon_{t-\Delta}B_t + \Delta C_t \\
or \quad \Delta C_t &=& S_t (\xi_t - \xi_{t-\Delta}) + B_t (\varepsilon_t - \varepsilon_{t-\Delta}) \no
\end{eqnarray}
which is different from $S_{t-\Delta}(\xi_t - \xi_{t-\Delta}) + B_{t-\Delta}(\varepsilon_t - \varepsilon_{t-\Delta})$. The above observation indicates that the external cash flow can be represented as a stochastic integral(but not in It\^{o} sense) resembling to $S_t d\xi_t + B_t d\varepsilon_t$. It would have the same integrator and integrand but should be defined by taking the right end points instead of left end points unlike the It\^{o} integral. However, here we confine ourselves in the formalism of It\^{o} calculus only. In order to derive an expression using It\^{o} integrals, we note that the equations \eqref{1a} and \eqref{eq.1.1} lead to the following discrete equation
\begin{eqnarray}
V_t - V_{t-\Delta} = \xi_{t-\Delta}(S_t - S_{t-\Delta}) + \varepsilon_{t-\Delta}(B_t - B_{t-\Delta}) + \Delta C_t \nonumber
\end{eqnarray}
or equivalently the SDE
\begin{eqnarray}\label{1f}
dV_t=\xi_t dS_t + \varepsilon_t dB_t + dC_t.
\end{eqnarray}
This observation essentially makes the following (see \cite{SAN} for details) definition, which is standard in the literature, self explanatory.
\noi\begin{defn1}
A strategy $\pi=(\xi,\varepsilon)$ is defined to be self financing if
\begin{eqnarray*}
dV_t = \xi_t dS_t + \varepsilon_t dB_t,\tab \forall t \geq 0.
\end{eqnarray*}
\end{defn1}
\noi Now using integration by parts rule of It\^{o} integration, we deduce from (\ref{1a})
\begin{eqnarray*}\label{1g}
dV_t=\xi_t dS_t + \varepsilon_t dB_t + S_t d\xi_t + B_td\varepsilon_t + d\langle S,\xi \rangle_t + d\langle B,\varepsilon\rangle_t.
\end{eqnarray*}
\noi By comparing this with equation (\ref{1f}) we get
\begin{eqnarray}\label{1c}
dC_t = S_t d\xi_t + B_td\varepsilon_t + d\langle S,\xi \rangle_t + d\langle B,\varepsilon\rangle_t.
\end{eqnarray}
\noi Since, $B_t$ is of finite variation and of continuous path, we derive
\begin{align*}
B_td\langle S^*,\xi\rangle_t= & d\langle BS^*,\xi\rangle_t+\xi_td\langle S^*,B\rangle_t-d\langle S^*\xi,B\rangle_t\\
=& d\langle S,\xi\rangle_t
\end{align*}
where $S^*_t:= B_t^{-1}S_t$. Thus using (\ref{1a}) and above identity, equation (\ref{1c}) gives
\begin{align*}
dC_t=&{S_td\xi_t}+B_t(dV_t^*-\xi_tdS^*_t-{S^*_td\xi_t}- {d\langle S^*,\xi\rangle_t})+B_t{d\langle S^*,\xi\rangle_t}\\
=& B_t(dV_t^*-\xi_t dS_t^*)
\end{align*}
or,
\begin{equation}\label{1d}
\frac{1}{B_t}dC_t = dV^*_t-\xi_tdS^*_t.
\end{equation}
\noi The process $C^*_t:=C^*_0 + \int^t_0 \frac{1}{B_t}dC_t$, for obvious reason, is called the \emph{discounted cost process} which gives the net
present value at $t=0$ of the accumulated additional cash flow up to time $t$. If a strategy $\pi$ is \emph{self-financing}, clearly $C^*_t(\pi)=$ constant and hence one has from \eqref{1d},
$$ dV^*_t = \xi_tdS^*_t .
$$
Often we encounter some market models where the class of self financing strategies is inadequate to ensure a perfect hedge for a given claim. Such markets are called \emph{incomplete}. In such a market an \emph{optimal strategy} is an admissible hedging strategy, need not be self financing, for which the \emph{quadratic residual risk}, a measure of the cash flow which would be specified in due course, is minimized subject to a certain constraint(see \cite{FS} for more details). It is shown in \cite{FS} that if the market is arbitrage free, the existence of an optimal strategy for hedging an $\calF_T$ measurable claim $H$, is equivalent to the existence of F\"{o}llmer Schweizer decomposition of discounted claim $H^*:= B^{-1}_T H$ in the form
\begin{equation}\label{eq1}
H^*=H_0+\int^{T}_{0}{\xi^{H^*}_t}dS^*_t+L^{H^*}_T
\end{equation}
where $H_0\in L^2(\Omega,\mathcal{F}_0,P), L^{H^*}=\{L^{H^*}_t\}_{0\leq t\leq T}$ is a square integrable martingale starting with zero and orthogonal to the martingale part of $S_t$, and $\xi^{H^*}=\{\xi^{H^*}_t\}$ satisfies A1 (i). Further $\xi^{H^*}$ appeared in the decomposition, constitutes the optimal strategy. Indeed the optimal strategy $\pi=(\xi_t,\varepsilon_t)$ is given by
\begin{eqnarray}\label{11a}
 \nonumber \xi_t &:=& \xi^{H^*}_t,\\
V^*_t &:=& H_0+\int^{t}_{0}{\xi_u}dS^*_u+L^{H^*}_t,\\
 \nonumber \varepsilon_t &:=& V^*_t-\xi_tS^*_t,
\end{eqnarray}
\noi and $B_t V^*_t$ represents the \emph{pseudo locally risk minimizing price} at time $t$ of the claim $H$. Hence the F\"{o}llmer Schweizer decomposition is the key thing to verify to settle the pricing and hedging problems in any given market (including incomplete).

\section{Market model and optimal hedging}
\subsection{Description of market model}\label{sec3.1}

Let $(\Omega,\mathcal{F},P)$
be the underlying complete probability space. Let the \emph{hypothetical state} of
the market be observable and modeled by $X= \{X_t\}_{t\geq 0}$, a semi-Markov process on a
finite state space $\mathcal{X}= \{ 1,2, \ldots,k\}$ with
transition probabilities $(p_{ij})$ and conditional holding time
distributions $F(\cdot\mid i)$. That is, if $0=T_0 <T_1<T_2< \cdots$
are time instances of consecutive transitions, then
$$P(X_{T_{n+1}}=j, T_{n+1}-T_n \leq y \mid X_{T_{n}}=i) = p_{ij} F(y\mid i).$$
We take the spot interest rate as the basis of locally risk free asset model and we assume that the
interest rate $r_t$ evolves over time depending on the state of the market. Apart from this locally risk free
money market account we further assume that the market consists only one stock as a risky asset.
Let $\{B_t\}_{t\geq0}$ be the price of money market account at time $t$ where, spot interest rate is $r_t=r(X_t)$ and $B_0=1$. We have $B_t = e^{\int_0^t r(X_{u}) du}$. Let $\{S_t\}_{t\geq 0}$ be the price process of the stock, which is governed by a semi-Markov modulated GBM i.e.,
\begin{equation}\label{8}
dS_t = S_t~(\mu(X_{t})dt +\sigma(X_{t}) dW_t),\tab S_0>0
\end{equation}where $ \{W_t\}_{t\geq 0}$ is a standard Wiener process
independent of $ \{X_t\}_{t\geq 0}$, $\mu : \mathcal{X} \to \mathbb{R}$
is the drift coefficient and $\sigma : \mathcal{X} \to (0, \infty)$
corresponds to the volatility. Let $\mathcal{F}_t$ be a filtration of $\mathcal{F}$ satisfying usual hypothesis
and right continuous version of the filtration generated by $X_t$ and $S_t$.
Clearly the solution of the above SDE is an $\mathcal{F}_t$ semimartingale with almost sure continuous paths. It is shown in \cite{AGMKG}
that this market model admits existence of an equivalent martingale measure. Hence under admissible strategy the market is
arbitrage free.

\noi Here the stock price is governed by two sources of uncertainties arising due to the driving Brownian motion, and the semi-Markov switching. The resulting market becomes incomplete hence the option pricing is rather involved. We know that in a \textit{complete market} every contingent claim can be replicated by a self-financing strategy but this is not the case in an incomplete market. To price a claim $H$ of European type in the above incomplete market, we would consider the pseudo locally risk minimizing pricing approach by F\"{o}llmer and Schweizer, i.e., decomposition of type \eqref{11a} and then show that the strategy, thus obtained is admissible.

\noi We represent the semi-Markov process $\{X_t\}$ as a stochastic integral with respect to a Poisson random measure which would play an important role later. We make the following assumptions which will be in effect throughout the paper.

\noi
\begin{itemize}
\item[(A2)](i) The transition matrix $(p_{ij})$
is irreducible.\\
(ii) For each $i$, $F(\cdot\mid i)$ has bounded and continuously differentiable derivative $f(\cdot\mid i)$.\\
(iii)$f(y\mid i)\neq 0$ for $y>0$.
\end{itemize}

\noi Embed $\mathcal{X}$ in $\mathbb{R}^k$ by identifying $i$ with
$e_i \in \mathbb{R}^k$. For $y \in [0,\infty)$, $i, j \in
\mathcal{X}$ let
\begin{eqnarray}\label{-1}
 \lambda_{ij}(y)&:=& p_{ij} \frac{f(y\mid i)}{1-F(y\mid i)} \geq
0\txt{ for } i\neq j,\\
\no \lambda_{ii}(y)&:=& -\sum_{j\in
\mathcal{X}, j\neq i} \lambda_{ij}(y) \txt{ for }i\in
\mathcal{X}.
\end{eqnarray}

\noi For $i\neq j\in \mathcal{X},~y\in \mathbb{R_+}$, let $\Lambda_{ij}(y)$ be the consecutive (with respect to the lexicographic ordering on $\mathcal{X}\times \mathcal{X}$) left closed and right open intervals of the real line, each having length $\lambda_{ij} (y)$. Define the functions $h: \mathcal{X}\times\mathbb{R}_+\times \mathbb{R} \to \mathbb{R}^k$ and $g: \mathcal{X} \times \mathbb{R}_+ \times \mathbb{R} \to \mathbb{R}_+ $ by
\begin{eqnarray} h(i,y,z)&:=&
\left\{ \begin{array}{l}\label{2a}
  j-i \tab\txt{ if } z\in \Lambda_{ij}(y) \\
  0 \tab\tab\txt{otherwise,}
\end{array} \right. \\
g(i,y,z)&:=& \left\{ \begin{array}{l}\label{2b}
  y \tab\tab\txt{if } z\in \Lambda_{ij}(y) \txt{ for some }
  j\neq i\\
  0 \tab\tab\txt{otherwise}.
\end{array}\right.
\end{eqnarray}
\noi We show in Theorem \ref{theo0} that $\{X_t\}_{t\geq0}$ can be described by the
following system of stochastic integral equations
\begin{eqnarray}\label{1}
  X_t &=& X_0 + \int_0^{t}\int_{\mathbb{R}}
h(X_{u-}, Y_{u-},z)\wp(du,dz)\\
\no Y_t &=& t- \int_0^{t} \int_{\mathbb{R}} g(X_{u-},
Y_{u-},z) \wp(du,dz)
\end{eqnarray}
where the integrations are over the interval $(0, t]$ and $\wp
(dt,dz)$ is the Poisson random measure with intensity $dtdz$, independent
of $X_0$.

\begin{theo}\label{theo0} The process $\{X_t\}$ defined in (\ref{1}) is a
semi-Markov process with transition probability matrix $(p_{ij})$
and conditional holding time distributions $F(y\mid i)$.
\end{theo}

\proof From (\ref{1}) it is clearly seen that $X_t$ is a right
continuous (since the integrations are over $(0,t]$) jump process
taking values in $\mathcal{X}$. Again from (\ref{2a}), (\ref{2b})
and (\ref{1}) for a fixed $\omega\in \Omega$, $\{X_t(\omega)\}$
has a jump at $t_0$ to a state $j$ if and only if
$\wp\left(\{t_0\} \times \Lambda_{X_{t_0-} (\omega)j}(Y_{t_0-}
(\omega))\right) (\omega)\neq 0$. By using this inductively for each
jump, we see that, $Y_{t_0}(\omega)=0$ if and only if
$\{X_t(\omega)\}$ has a jump at $t_0$. Let $T_n$ denote the time
of $n$th jump of $X_t$, whereas $T_0:= 0$ and $\tau_n:=T_n-T_{n-1}$. For
a fixed $t$, let $n(t):= \max\{n: T_n \leq t \}$. Thus
$T_{n(t)}\leq t < T_{n(t)+1}$ and $Y_t = t-T_{n(t)}$. Hence,
using the property of Poison random measure
\begin{eqnarray*}
P\left(\txt{ No jump in }(T_n, T_n+y ] \mid \mathcal{F}_{T_n}\right)&=&P \left(\wp \left(\bigcup_{0< s\leq y}\left(\{T_n+s\}\times \bigcup_{j\neq X_{T_n}}\Lambda_{X_{T_n}j}(s)\right)\right)=0 \right)\\
&=&\exp\left( -\int_0^y \sum_{j\neq X_{T_n}} \lambda_{X_{T_n}j}(s)ds \right).
\end{eqnarray*}
Again, from (\ref{-1}), we have
\begin{eqnarray*}
\frac{d}{dy}F(y\mid i) = (1-F(y\mid i)) \sum_{j\neq i}\lambda_{ij}(y)\tab \mathrm{for}~ \mathrm{any}~y>0,~i\in \chi;
\end{eqnarray*}
which gives $$\exp\Big(\ds{ -\int_0^y }\sum_{j\neq i}
\lambda_{ij}(s)ds \Big) = 1-F(y\mid i).$$ Thus
\begin{equation}\label{3c}
P(\tau_{n+1} \le y \mid \mathcal{F}_{T_n}) = F( y \mid X_{T_n}).
\end{equation}
\noi Then using Markovity of $\{(X_t,Y_t)\}_{t\geq0}$
\begin{eqnarray}\label{3d}
\no P(X_{T_{n+1}}=j\mid \mathcal{F}_{T_{n+1}-})&=&P(X_{T_{n+1}}=j\mid X_{T_{n+1}-}=X_{T_{n}},Y_{T_{n+1}-}=T_{n+1}-T_{n})\\
\no &=&P[\int_{\mathbb{R}}h(X_{T_{n}},T_{n+1}-T_{n},z)\wp(\{T_{n+1}-T_{n}\}\times dz)=j-X_{T_{n}}\mid \\
\no &&\int h(X_{T_{n}},T_{n+1}-T_{n},z)\wp(\{T_{n+1}-T_{n}\}\times dz)\neq 0]\\
\no &=&P[\wp(\{T_{n+1}-T_{n}\}\times\Lambda_{X_{T_{n}}j}(T_{n+1}-T_{n})\neq 0 \mid \wp(\{T_{n+1}-T_{n}\}\\
\no &&\times\Lambda_{X_{T_{n}}j}(T_{n+1}-T_{n})\neq 0~\textrm{for}~\textrm{some}~j)]\\
\no &=&\frac{\mid \Lambda_{X_{T_{n}}j}(T_{n+1}-T_{n})\mid}{\mid \cup_{j\neq i}\Lambda_{X_{T_{n}}j}(T_{n+1}-T_{n})\mid}\\
\no &=&\frac{\lambda_{X_{T_{n}}j}(T_{n+1}-T_{n})}{\sum_{j\neq X_{T_{n}}} \lambda_{X_{T_{n}}j}(T_{n+1}-T_{n})}\\
\no &=&p_{X_{T_n}j}\\
   &=&P(X_{T_{n+1}} =j \mid X_{T_{n}}).
\end{eqnarray}
We also have
\begin{eqnarray*}
\no P(X_{T_{n+1}}=j, \tau_{n+1} \le y \mid \mathcal{F}_{T_{n}})
    &=& \int_0^y\exp\Big(\ds{ -\int_0^u }\sum_{j\neq X_{T_{n}}} \lambda_{X_{T_{n}}j}(t)dt \Big) \lambda_{X_{T_{n}}j}(u) du\\
\no &=& \int_0^y (1-F(u\mid X_{T_{n}}))p_{X_{T_{n}}j}\frac{f(u\mid X_{T_{n}})}{1-F(u\mid X_{T_{n}})} du\\
 &=& p_{X_{T_{n}}j} F(y\mid X_{T_{n}}).
\end{eqnarray*}
The above equation along with \eqref{3c} and \eqref{3d} proves the theorem.\qed
\subsection{B-S-M type equation for European call price}
Now onward we consider a particular contingent claim i.e., a European call option
on $\{S_t\}$ with strike price $K$ and maturity time $T$. In this case
the $\calF_T$ measurable contingent claim $H$ is given by
\begin{equation}\label{4}
H=(S_T-K)^+.
\end{equation}

\noi To find an optimal hedging strategy for this claim in the semi-Markov modulated
market, we consider the following system of (integro-partial)
differential equations given by
\begin{eqnarray}\label{p1}
\no \frac{\partial}{\partial t} \varphi(t, s, i, y)+
\frac{\partial} {\partial y} \varphi(t, s, i, y) + r( i ) s
\frac{\partial} {\partial s} \varphi(t, s, i, y) + \frac{1}{2}
\si^2( i ) s^2 \frac{\partial^2} {\partial s^2} \varphi(t, s, i, y)\\
+\frac{f(y\mid i)}{1-F(y\mid i)}\sum_{j\neq i}p_{ij}[\varphi(t,
s, j, 0) -\varphi(t, s, i, y)] = r( i ) ~ \varphi(t, s, i, y),
\end{eqnarray}
\noi defined on
\begin{equation}\label{D}
\mathcal{D}:= \{ (t, s, i, y)\in (0,T)\times \mathbb{R_+}\times
\mathcal{X}\times (0,T) \mid y \in (0,t)\},
\end{equation}
\noi and with conditions
\begin{eqnarray}\label{18}
\no \varphi (t, 0, i, y)&=& 0 \tab \forall t \in [0,T]\\
\varphi (T, s, i, y)&=& (s-K)^+; \tab s\in \mathbb{R_+}; \tab 0 \le
y\le T ; \tab i= 1, 2, \ldots, k
\end{eqnarray}
\noi where $r(\cdot)$, $\sigma(\cdot)$, $(p_{ij})$, $f(\cdot\mid i)$ and $F(\cdot\mid i)$ are as in Section 3.1. We would show,
in the next subsection that the solution of \eqref{p1}-\eqref{18} gives the locally risk minimizing price function of the European
call option \eqref{4}. But before that, it remains to establish the existence and uniqueness of solution of the above non local differential equation. This we accomplish in two steps. First in the following lemma we consider a Volterra integral equation of second kind and establish existence and uniqueness result of that. Then we show in couple of Propositions, that the PDE and the IE problems are ``equivalent". Thus we obtain the existence and uniqueness of the PDE in Theorem \ref{theo1} at the end of this subsection.
\begin{lem}\label{lm2}
Consider the following integral equation
\begin{eqnarray}
\no\varphi(t,s,i,y)&=&\frac{1- F(T-t+y\mid i)}{1-F(y\mid i)}
\eta_{i}(t,s)+\int_0^{T-t} e^{-r(i)v}
\frac{f(y+v\mid i)} {1-F(y\mid i)} \times\\
&& \sum_j p_{i j} \int_0^{\infty} \varphi(t+v,x,j,0)
\frac{e^{\frac{-1}{2}((\ln(\frac{x}{s})-(r(i)
-\frac{\si^2(i)}{2})v) \frac{1}{\si(i)\sqrt v})^2}}{\sqrt{2\pi}
x\si(i)\sqrt v} dx dv \label{34}\\
\label{35} \txt{ with }\varphi(t,0,i,y)&=&0 ~\forall t \in [0,T], i\in \chi,~y\in [0,t]
\end{eqnarray}
where $\eta_{i}(t,s)$ is the standard Black-Scholes price of the same European call option with fixed interest rate $r(i)$ and volatility $\sigma(i)$. Then
\begin{enumerate}[i.]
  \item the problem \eqref{34}-\eqref{35} has unique solution with at most linear growth in $s$ variable and the solution is non-negative.
  \item the solution of the integral equation is in $C^{1,2,1}(\mathcal{D})$.
\end{enumerate}
\end{lem}
\proof (i) We first note that a solution of \eqref{34}-\eqref{35} is a fixed point of the operator $A$ and vice versa, where
\begin{eqnarray}
\no A \varphi(t,s,i,y)&:=& \frac{1- F(T-t+y\mid i)}{1-F(y\mid i)}\eta_{i}(t,s)+\int_0^{T-t}e^{-r(i) v} \frac{f(y+v\mid i)}{1-F(y\mid i)}
 \sum_j p_{ij}\\
 \no && \int^\infty _0 \varphi(t+v,x,j,0) \alpha(x;s,i,v)dx dv,
\end{eqnarray}
and $\alpha(x;s,i,v):= \frac{e^{\frac{-1}{2}((\ln(\frac{x}{s})-(r(i)
-\frac{\si^2(i)}{2})v) \frac{1}{\si(i)\sqrt
v})^2}}{\sqrt{2\pi} x\si(i)\sqrt v} $, as a function of $x$, is the log normal probability density function. Set
$$B=\left\{\varphi:\mathcal{\overline{D}}\rightarrow [0,\infty), \mathrm{continuous} \mid \varphi(\cdot,0,\cdot,\cdot)=0, ~\|\varphi\|:=\sup_{\mathcal{D}}\mid\frac{\varphi(t,s,i,y)}{1+s}\mid < \infty\right\}.
$$
It is easy to check that $B$ is a closed subset of a Banach space $(\mathcal{B}, \|~ \|)$, where $\mathcal{B}$ is the set of all continuous functions with at most linear growth in $s$ variable. Now in order to show existence and uniqueness in the prescribed class, it is sufficient to show that $A$ is a contraction. Because, then Banach fixed point theorem ensures existence and uniqueness of the fixed point. To show that $A$ is a contraction, we need to show $||A\varphi_1-A\varphi_2|| \leq L||\varphi_1-\varphi_2||$ where $L<1$. Indeed
\begin{eqnarray*}
\|A\varphi_1-A\varphi_2\|&=&\sup_{\mathcal{D}}\bigg|\frac{ A\varphi_1-A\varphi_2}{1+s}\bigg|\\
&=&\sup_{\mathcal{D}}\bigg|\int^{T-t}_0 e^{-r(i) v} \frac{f(y+v\mid i)}{1-F(y\mid i)} \sum p_{ij} \int^\infty_0 (\varphi_1-\varphi_2)(t+v,x,j,0)\frac{\alpha(x;s,i,v)}{1+s}dx dv\bigg|\\
&\leq& \sup_{\mathcal{D}} \bigg| \int^{T-t}_0 e^{-r(i) v} \frac{f(y+v\mid i)}{1-F(y\mid i)}\sum p_{ij} \int^\infty_0 (1+x)~\sup_{\mathcal{D}}\bigg|\frac{\varphi_1-\varphi_2}{1+x}\bigg| \frac{\alpha(x;s,i,v)}{1+s}dx dv\bigg|\\
&=& \sup_\mathcal{D} \bigg| \int^{T-t}_0 e^{-r(i) v} \frac{f(y+v \mid i)}{1-F(y \mid i)}\|\varphi_1- \varphi_2 \| \frac{a(s)}{1+s}dv\bigg|
\end{eqnarray*}
where,
\begin{eqnarray*}
a(s)&=& \int^\infty_0 (1+x) \alpha(x;s,i,v) dx\\
&=& 1 + e^{\ln s + \left( r(i)-\frac{\sigma^2(i)}{2}\right)v+\frac{\sigma^2(i) v}{2}}\\
&=& 1+e^{\ln s+r(i)v}\\
&=& 1+ s e^{r(i)v}.
\end{eqnarray*}
Thus, $\|A\varphi_1-A\varphi_2\| \, \leq\, L\|\varphi_1-\varphi_2\|$ where,
\begin{eqnarray*}
L&=&\sup_\mathcal{D} \bigg|\int^{T-t}_0e^{-r(i) v}\frac{f(y+v \mid i)}{1-F(y \mid i)}\frac{1+se^{r(i)v}}{1+s}dv\bigg|\\
&=&\sup_\mathcal{D}\bigg( \frac{1}{1-F(y\mid i)}\bigg|\int^{T-t}_0 f(y+v|i)\frac{e^{-r(i)v}+s}{{1+s}}dv\bigg|\bigg)\\
&\leq &\sup_\mathcal{D}\bigg( \frac{1}{1-F(y\mid i)}\int^{T-t}_0 f(y+v|i)dv\bigg)\\
&=&\sup_\mathcal{D}\bigg(\frac{F\left( y+T-t\mid i\right) -F(y|i)}{1-F\left(y|i\right)}\bigg)\\
&<&\frac{1-F(y|i)}{1-F(y|i)}\\
&=&1
\end{eqnarray*}
using $r(i)\geq 0$ and (A2).

\noi (ii) Now we would establish the desired regularity. Using (A2) and smoothness of $\eta_i$ for each $i$, the first term on the right hand side is in $C^{1,2,1}(\mathcal{D})$. Under assumption (A2) the fact, the second
term is continuous differentiable in $y$ and twice continuously differentiable in s, follows immediately. The continuous
differentiability in $t$ follows from the fact that the term $\varphi(t+v,x,j,0)$
is multiplied by $C^1((0,\infty))$ functions in $v$ and then integrated over $v\in(0,T-t)$.
Hence $\varphi(t,s,i,y)$ is in $C^{1,2,1}(\mathcal{D})$. Finally the continuity of
$\varphi$ on $\mathcal{\bar{\mathcal{D}}}$ follows trivially. \qed

\begin{prop}\label{theo3} The unique solution of \eqref{34}-\eqref{35} also solves the initial boundary value problem \eqref{p1}-\eqref{18}.
\end{prop}
\proof
We consider the solutions of \eqref{34}-\eqref{35}. Then we have
\begin{eqnarray}\label{1ae}
\no\varphi(t,s,i,y)&=&\frac{1- F(T-t+y\mid i)}{1-F(y\mid i)} \eta_{i}(t,s)+\int_0^{T-t} e^{-r(i)v}
\frac{f(y+v\mid i)} {1-F(y\mid i)} \times\\
&& \sum_j p_{i j} \int_0^{\infty} \varphi(t+v,x,j,0)\alpha(x;s,i,v) dx dv
\end{eqnarray}
where
\begin{equation*}
\alpha(x;i,s,v)=\frac{e^{-\frac{1}{2}L(i)^2}}{\sqrt{2\pi}x\sigma(i)\sqrt{v}},~~~~
L(i)=\frac{\ln\left(\frac{x}{s}\right)-\left(r(i)-\frac{\si^2(i)}{2}\right)v}{\sigma(i)\sqrt{v}}
\end{equation*}
and $\eta_{i}(t,s)$ is the price of the call option under standard B-S-M assumption with constant regime $i$. Hence
\begin{equation}\label{eq2}
\frac{\partial\eta_{i}(t,s)}{\partial t} + r(i)s\frac{\partial\eta_{i}(t,s)}{\partial s}+\frac{1}{2}\sigma^2(i)s^2 \frac{\partial^2\eta_{i}(t,s)}{\partial s^2} = r(i)\eta_{i}(t,s)
\end{equation}
and $\eta_{i}(T,s)=(s-K)^+$. Thus using \eqref{1ae}, $\varphi(T,s,i,y)=\eta_{i}(T,s)=(s-K)^+$, i.e., the condition \eqref{18} holds. Also, by a direct calculation one has
\begin{equation}\label{10ae}
L(i)\frac{\partial L(i)}{\partial v}+r(i)\frac{L(i)}{\sigma(i) \sqrt{v}}+\frac{1}{2}\frac{L(i)^2}{v}-\frac{\sigma (i)L(i)}{2\sqrt{v}}=0.
\end{equation}
From Lemma \ref{lm2} (ii), $\varphi$ is in $C^{1,2,1}(\mathcal{D})$. Hence we can perform the partial differentiations w.r.t. $t$ and $y$ on the both sides of \eqref{1ae}. We obtain
\begin{eqnarray}\label{2ae}
\no \lefteqn{\frac{\partial}{\partial t} \varphi(t, s, i, y)}\\
\no&=&\frac{f(T-t+y|i)}{1-F(y\mid i)}\eta_{i}(t,s)+\frac{ 1- F(T-t+y\mid i)}{\left(1-F(y\mid i)\right)}\frac{\partial\eta_{i}(t,s)}{\partial t}
- e^{-r(i)(T-t)}\frac{f(T-t+y\mid i)}{1-F(y\mid i)}\\
&&\no\sum p_{ij}\int_0^\infty \varphi(T,x,j,0)\alpha(x;s,i,T-t)dx+\int^{T-t}_0 e^{-r(i)v}\frac{f(y+v\mid i)}{1-F(y\mid i)} \\
&&\sum p_{ij}\int^\infty_0 \frac{\partial \varphi}{\partial t}(t+v,x,j,0) \alpha(x;s,i,v) dx dv
\end{eqnarray}
by differentiating w.r.t. $t$ under the sign of integral. Now we operate partial derivative w.r.t. $y$ on both sides of \eqref{1ae} after simplifying the right side using integration by parts w.r.t. $v$ where $f(y+v\mid i)$ is treated as second function to get
\begin{eqnarray}\label{3ae}
\no \lefteqn{\frac{\partial}{\partial y} \varphi(t, s, i, y)}\\
&=&\no -\frac{f(T-t+y \mid i)}{1-F(y\mid i)}\eta_{i}(t,s)+\frac{1- F(T-t+y\mid i)}{\left(1-F(y\mid i)\right)^2}f(y|i)\eta_{i}(t,s)\\
&&\no +\frac{f(y|i)}{1-F(y\mid i)}\Big(\varphi(t,s,i,y)- \frac{1- F(T-t+y \mid i)}{1-F(y\mid i)}\eta_{i}(t,s)\Big)+e^{-r(i)(T-t)}\frac{f(T-t+y\mid i)}{1-F(y\mid i)} \\
&&\no\sum p_{ij} \int^\infty_0 \varphi(T,x,j,0)\alpha(x;s,i,T-t)dx-\frac{f(y\mid i)}{1-F(y\mid i)}\sum p_{ij}\varphi(t,s,j,0)\\
&&\no-\int^{T-t}_0 e^{-r(i) v} \frac{f(y+v\mid i)}{1-F(y\mid i)}\int^\infty_0 \alpha(x;s,i,v) \Bigg\{-r(i) \sum p_{ij} \varphi(t+v,x,i,0)\\
&&-\sum p_{ij} \varphi(t+v,x,j,0)\left(L(i) \frac{\partial L(i)}{\partial v}+\frac{1}{2v}\right)+\sum p_{ij} \frac{\partial{\varphi(t+v,x,j,0)}}{\partial t}\Bigg\}dx dv.
\end{eqnarray}
\noi By adding equations \eqref{2ae} and \eqref{3ae}, we get
\begin{eqnarray}\label{6ae}
\no\lefteqn{\frac{\partial}{\partial t} \varphi(t, s, i, y)+\frac{\partial} {\partial y} \varphi(t, s, i, y)}\\
&=&\no\frac{ 1- F(T-t+y\mid i)}{1-F(y\mid i)}\frac{\partial\eta_{i}(t,s)}{\partial t}+ \frac{f(y|i)}{1-F(y\mid i)}\left(
\varphi(t,s,i,y)-\sum p_{ij}\varphi(t,s,j,0)\right) + \int^{T-t}_0 e^{-r(i)v}\\
&& \frac{f(y+v|i)}{1-F(y|i)}\sum p_{ij} \int_0^\infty \varphi(t+v,x,j,0)\alpha(x;s,i,v)\left(r(i)+L(i)\frac{\partial L(i)}{\partial v}+\frac{1}{2v}\right)dv dx.
\end{eqnarray}
\noi Now we differentiate both sides of \eqref{1ae} w.r.t. $s$ once and twice respectively and obtain
\begin{eqnarray}\label{4ae}
\no \lefteqn{\frac{\partial}{\partial s} \varphi(t, s, i, y)} \\
\no&=& \frac{ 1- F(T-t+y\mid i)}{1-F(y\mid i)}\frac{\partial\eta_{i}(t,s)}{\partial s}+\int_0^{T-t} e^{-r(i)v}\frac{f(y+v\mid i)} {1-F(y\mid i)} \sum_j p_{i j}\int_0^{\infty} \varphi(t+v,x,j,0)\\
&&\alpha(x;s,i,v)\frac{L(i)}{\sigma(i)\sqrt{v} s} dx dv,
\end{eqnarray}
\begin{eqnarray}\label{5ae}
\no\lefteqn{\frac{\partial^2} {\partial s^2} \varphi(t, s, i, y)}\\
\no&=& \frac{ 1- F(T-t+y\mid i)}{1-F(y\mid i)}\frac{\partial^2 \eta_{i}(t,s)}{\partial s^2}+\int_0^{T-t} e^{-r(i)v}\frac{f(y+v\mid i)} {1-F(y\mid i)}
\sum_j p_{i j} \int_0^{\infty} \varphi(t+v,x,j,0)\\
&&\alpha(x;s,i,v) \frac{1}{s^2}\left(\frac{L^2(i)}{\sigma^2(i) v} - \frac{L(i)}{\sigma(i)\sqrt{v}} -\frac{1}{\sigma^2(i) v}\right)dx dv.
\end{eqnarray}
\noi From equations \eqref{4ae} and \eqref{5ae}, we get
\begin{eqnarray}\label{7ae}
\no \lefteqn{r(i) s \frac{\partial \varphi}{\partial s}+ \frac{1}{2}\sigma^2(i) s^2 \frac{\partial^2 \varphi}{\partial s^2}}\\
&=&\no \frac{ 1- F(T-t+y\mid i)}{1-F(y\mid i)}\left[r(i)s\frac{\partial\eta_{i}(t,s)}{\partial s}+\frac{1}{2}\sigma^2(i)s^2 \frac{\partial^2\eta_{i}(t,s)}{\partial s^2}\right]+\int_0^{T-t}e^{-r(i)v}\frac{f(y+v\mid i)}{1-F(y|i)}\\
&&\sum p_{ij}\int_0^\infty \varphi(t+v,x,j,0)\alpha(x;s,i,v)\left(\frac{r(i)L(i)}{\sigma(i)\sqrt{v}}+\frac{L^2(i)}{2v}-\frac{\sigma(i)}{2\sqrt{v}}L(i)-\frac{1}{2v}\right)dx dv.
\end{eqnarray}
\noi Finally, from equations \eqref{1ae}, \eqref{eq2}, \eqref{10ae}, \eqref{6ae} and \eqref{7ae} we get
\begin{eqnarray*}\label{8ae}
\lefteqn{\no\frac{\partial}{\partial t} \varphi(t, s, i, y)+\frac{\partial} {\partial y} \varphi(t, s, i, y)+r(i) s \frac{\partial }{\partial s}\varphi(t, s, i, y)+ \frac{1}{2}\sigma(i)^2(i) s^2\frac{\partial^2 }{\partial s^2}\varphi(t, s, i, y)}\\
&=&\no \frac{ 1- F(T-t+y\mid i)}{1-F(y\mid i)}\left[\frac{\partial\eta_{i}(t,s)}{\partial t} + r(i)s\frac{\partial\eta_{i}(t,s)}{\partial s}+\frac{1}{2}\sigma^2(i)s^2 \frac{\partial^2\eta_{i}(t,s)}{\partial s^2}\right] -\frac{f(y\mid i)}{1-F(y\mid i)}\times \\
&&\sum_{j\neq i} p_{ij}(\varphi(t,s,j,0)-\varphi(t,s,i,y)) + r(i) \left(\varphi(t,s,i,y)- \frac{ 1- F(T-t+y\mid i)}{1-F(y\mid i)} \eta_{i}(t,s)\right)\\
&=& -\frac{f(y\mid i)}{1-F(y\mid i)}\sum_{j\neq i} p_{ij}(\varphi(t,s,j,0)-\varphi(t,s,i,y))+r(i)\varphi(t,s,i,y).
\end{eqnarray*}
\noi Thus
\begin{eqnarray*}\label{9ae}
\no \frac{\partial}{\partial t} \varphi(t, s, i, y)+
\frac{\partial} {\partial y} \varphi(t, s, i, y) + r( i ) s
\frac{\partial} {\partial s} \varphi(t, s, i, y) + \frac{1}{2}
\si^2( i ) s^2 \frac{\partial^2} {\partial s^2} \varphi(t, s, i, y)\\
+\frac{f(y\mid i)}{1-F(y\mid i)}\sum_{j\neq i}p_{ij}[\varphi(t,
s, j, 0) -\varphi(t, s, i, y)] = r( i ) ~ \varphi(t, s, i, y). \qed
\end{eqnarray*}

\begin{prop}\label{theo4} Let $\varphi$ be a classical solution of \eqref{p1}-\eqref{18}. Then $\varphi$ also solves the integral equation \eqref{34}-\eqref{35}.
\end{prop}
\proof Let $(\tilde{\Omega},\tilde{\mathcal{F}},\tilde P)$ be a probability space which holds a standard Brownian motion $\tilde{W}$ and a semi-Markov process $\tilde{X}$ independent of
$\tilde{W}$ such that the transition probabilities and holding time distribution of $\tilde{X}$ are as same as of $X$. Let $\tilde{B}_t$ and $\tilde{S}_t$ be given by
\begin{eqnarray}\label{37}
\no \tilde{B}_t&=&e^{\int_0^t r(\tilde{X}_u)du},\\
d\tilde{S}_t&=& \tilde{S}_t(r(\tilde{X}_t)dt + \si(\tilde{X}_t)
d\tilde{W}_t),\tab \tilde{S}_0>0.
\end{eqnarray}
Let $\tilde{Y}_t$ represent the amount of time the process $\tilde{X}_t$ is at the current state since the last jump and $ \tilde{\mathcal{F}}_t$ be the underlying filtration satisfying the usual hypothesis. Thus $\tilde{P}$ is a risk-neutral measure for the risky asset $\tilde{S}$, given by \eqref{37}. Let the consecutive jump times be $0=T_0<T_1<T_2<\cdots$ and $n(t) := \max\{n\ge 0 \mid T_n \le t\}$. Hence, $T_{n(t)}=t- \tilde{Y}_t$. We observe that the process $(\tilde{S}_t,\tilde{X}_t,\tilde{Y}_t)$ is jointly Markov with infinitesimal generator $\tilde{\mathcal{A}}$ given by
\begin{eqnarray}\label{gen}
\no \tilde{\mathcal{A}} \varphi(s,i,y)=\frac{\partial}{\partial y}
\varphi (s, i, y) + r(i) s \frac{\partial}{\partial s} \varphi (s,
i, y)+ \frac{1}{2} \si^2(i) s^2 \frac{\partial^2} {\partial
s^2} \varphi (s, i, y)\\
+~\frac{f(y\mid i)}{1-F(y\mid i)}\sum_{j\neq i}p_{ij}
[\varphi(s,j,0) -\varphi(s,i,y)]
\end{eqnarray}
for every function $\varphi$ which is twice differentiable in $s$ and once differentiable in $y$. Let $\vf$ be a classical solution of \eqref{p1}-\eqref{18}. Therefore, from \eqref{p1} one has $\frac{\partial \varphi}{\partial t }(t,s,i,y) + \tilde{\mathcal{A}} \varphi (t,s,i,y)= r(i) \varphi (t,s,i,y)$. Now by using this and the It\^{o}'s formula on $N_t := e^{-\int_0^t r(\tilde{X}_u)du} \varphi(t,\tilde{S}_t,\tilde{X}_t,\tilde{Y}_t)$ under the measure $\tilde{P}$, we get
\begin{eqnarray*}
  dN_t &=& e^{-\int_0^t r(\tilde{X}_u)du}\left(- r(\tilde{X}_t) \varphi (t,\tilde{S}_t,\tilde{X}_t,\tilde{Y}_t)+ \frac{\partial \varphi}{\partial t }(t,\tilde{S}_t,\tilde{X}_t,\tilde{Y}_t) + \tilde{\mathcal{A}} \varphi (t,\tilde{S}_t,\tilde{X}_t,\tilde{Y}_t)\right)dt + d\tilde{M}_t \\
   &=& d\tilde{M}_t
\end{eqnarray*}
where $\tilde{M}_t$ is a martingale. Hence $N_t$ is an $ \mathcal{\tilde F}_t$ martingale under $ {\tilde P}$. Thus
\begin{eqnarray}\label{aa}
\no\varphi(t,\tilde{S}_t,\tilde{X}_t,\tilde{Y}_t)&=& e^{\int_0^t r(\tilde{X}_u)du}N_t\\
\no&=& \tilde{E}[e^{\int_0^t r(\tilde{X}_u)du}N_T\mid \mathcal{\tilde F}_t]\\
&=& \tilde{E}[e^{-\int_t^T r(\tilde{X}_u)du} (\tilde{S}_T-K)^+\mid \mathcal{\tilde F}_t].
\end{eqnarray}
Now by conditioning at transition times and using the conditional lognormal distribution of stock price process, we have
\begin{eqnarray*}
\lefteqn{\varphi(t,\tilde{S}_t,\tilde{X}_t,\tilde{Y}_t)}\\
&=& \tilde{E}[e^{-\int_t^T r(\tilde{X}_u)du} (\tilde{S}_T-K)^+\mid
\tilde{S}_t, \tilde{X}_t, \tilde{Y}_t]\\
&=& \tilde{E}[\tilde{E}[e^{-\int_t^T r(\tilde{X}_u)du}
(\tilde{S}_T-K)^+ \mid \tilde{S}_t, \tilde{X}_t, \tilde{Y}_t,
\tilde{T}_{n(t)+1}]\mid \tilde{S}_t, \tilde{X}_t, \tilde{Y}_t]\\
&=& P(\tilde{T}_{n(t)+1} > T\mid \tilde{X}_t,\tilde{Y}_t) \tilde{E}[e^{-\int_t^T
r(\tilde{X}_u)du} (\tilde{S}_T-K)^+ \mid \tilde{S}_t, \tilde{X}_t,
\tilde{Y}_t,\tilde{T}_{n(t)+1} > T] \\
&& + \int_0^{T-t} \tilde{E}[e^{-\int_t^T r(\tilde{X}_u)du}
(\tilde{S}_T-K)^+\mid \tilde{S}_t, \tilde{X}_t, \tilde{Y}_t,
\tilde{T}_{n(t)+1}=t + v] \frac{f(t-T_{n(t)}\mid
\tilde{X}_t)} {1-F(\tilde{Y}_t\mid \tilde{X}_t)}dv\\
&=& \frac{ 1- F(T-T_{n(t)}\mid
\tilde{X}_t)}{1-F(\tilde{Y}_t\mid \tilde{X}_t)}
\eta_{\tilde{X}_t}(t,\tilde{S}_t)+ \int_0^{T-t}
e^{-r(\tilde{X}_t)v} \frac{f(\tilde{Y}_t+v\mid \tilde{X}_t)}
{1-F(\tilde{Y}_t\mid \tilde{X}_t)} \times \\
&& \sum_j p_{\tilde{X}_tj} \int_0^{\infty}
\tilde{E}[e^{-\int_{t+v}^T r(\tilde{X}_u)du}
(\tilde{S}_T-K)^+\mid \tilde{S}_{t+v}=x,
\tilde{S}_t,\tilde{Y}_{t+v}=0,\\
&& \tilde{X}_{t+v}=j, \tilde{T}_{n(t)+1}=t+v]
\frac{e^{\frac{-1}{2}((\ln(\frac{x}{\tilde{S}_t})-(r(\tilde{X}_t)
-\frac{\si^2(\tilde{X}_t)}{2})v) \frac{1}{\si(\tilde{X}_t)\sqrt
v})^2}}{\sqrt{2\pi}\si(\tilde{X}_t)\sqrt{v} x} dx dv\\
&=& \frac{ 1- F(T-t+\tilde{Y}_t\mid
\tilde{X}_t)}{1-F(\tilde{Y}_t\mid \tilde{X}_t)}
\eta_{\tilde{X}_t}(t,\tilde{S}_t)+\int_0^{T-t}
e^{-r(\tilde{X}_t)v}
\frac{f(\tilde{Y}_t+v\mid \tilde{X}_t)}
{1-F(\tilde{Y}_t\mid \tilde{X}_t)} \times\\
&& \sum_j p_{\tilde{X}_t j} \int_0^{\infty} \varphi(t+v,x,j,0)
\frac{e^{\frac{-1}{2}((\ln(\frac{x}{\tilde{S}_t})-(r(\tilde{X}_t)
-\frac{\si^2(\tilde{X}_t)}{2})v) \frac{1}{\si(\tilde{X}_t)\sqrt
v})^2}}{\sqrt{2\pi} x\si(\tilde{X}_t)\sqrt v} dx dv.
\end{eqnarray*}
where $\eta_{i}(t,s)$ is the standard Black-Scholes price of European call option with fixed interest rate $r(i)$ and volatility $\sigma(i)$. Finally by using irreducibility condition (A2), we can replace $(\tilde S_t,\tilde X_t,\tilde Y_t)$ by generic variable $(s,i,y)$ in the above relation and thus conclude that $\varphi$ is a solution of \eqref{34}-\eqref{35}. \qed
\begin{theo}\label{theo1}
The initial-boundary value problem \eqref{p1}-\eqref{18} has a unique classical solution in the class of functions with at most linear growth.
\end{theo}
\proof Existence follows from Lemma \ref{lm2} and Proposition \ref{theo3}. For uniqueness, first assume that $\vf_1$ and $\vf_2$ are two classical solutions of \eqref{p1}-\eqref{18} in the prescribed class. Then using Proposition \ref{theo4}, we know that both also solve \eqref{34}-\eqref{35}. But from Lemma \ref{lm2}, there is only one such in the prescribed class. Hence $\vf_1=\vf_2$.\qed
\begin{rem}
A different proof of the above theorem appears in \cite{AGMKG} which does not require the detailed study of the Volterra integral equation. Nevertheless, that proof heavily depends on the mild solution techniques \cite{PA} and Proposition 3.1.2 of \cite{CFMW}. On the other hand, the present proof is self contained and needs no results from the theory of parabolic PDEs.
\end{rem}
\subsection{Optimal hedging}
\begin{theo}\label{theo5} Let $\vf$ be the unique classical solution of \eqref{p1}-\eqref{18} in the class of functions with at most linear growth.
\begin{enumerate}[i.]
  \item Let $(\xi,\varepsilon)$ be given by
  \begin{equation}\label{VI3.20}
 \xi_t :=\frac{\partial\varphi(t,S_t,X_{t-},Y_{t-})}{\partial s} \txt{ and } \varepsilon_{t} := e^{-\int_{0}^{t}r(X_{u})du} (\varphi(t,S_t,X_{t},Y_{t})-\xi_{t}S_{t}).
 \end{equation}
Then $(\xi,\varepsilon)$ is the optimal admissible strategy.
  \item $\varphi(t,S_t,X_t,Y_t)$ is the locally risk minimizing price of $(S_T-K)^+$.
\end{enumerate}
\end{theo}

\proof (i) Under the market model give in Subsection \ref{sec3.1}, the mean variance tradeoff (MVT) process $\hat{K}_t$ (as defined in Pham et al \cite{PH}) takes the following form
\begin{equation*}
\hat{K}_t=\int_0^t\left(\frac{\mu(X_s)-r(X_s)}{\sigma(X_s)}\right)^2 ds.
\end{equation*}
Hence $\hat{K}_t$ is bounded and continuous on $[0,T]$. We also know that $S_t$ has almost sure continuous paths. Since, $H^*\in L^2(\Omega, {\cal F}, P)$ for $H=(S_T-K)^+$ we apply corollary 5 and Lemma 6 of \cite{PH} to conclude that $H^*$ admits F-S decomposition \eqref{eq1} with an integrand $\xi^{H^*}$ satisfying A1 (i) and $L^{H^*}$ being square integrable. Therefore, to prove the theorem it is sufficient to show that
\begin{enumerate}[(a)]
  \item there exists $\mathcal{F}_0$ measurable $H_0$ and $\mathcal{F}_T$ measurable $L_T$ such that $L_t:= E[L_T\mid \mathcal{F}_t]$ is orthogonal to $\int_0^t \sig(X_t)S^*_t dW_t$ i.e., the martingale part of $S^*_t$ and $H^*=H_0+\int^{T}_{0}{\xi_t}dS^*_t+L_T$;
  \item $\frac{1}{B_t} \varphi(t,S_t,X_{t-},Y_{t-})= H_0+\int^{t}_{0}{\xi_t}dS^*_t+L_t$ for all $t\le T$;
  \item $\varphi(t,S_t,X_{t},Y_{t})= B_t \varepsilon_{t} + \xi_{t}S_{t}$ for all $t\le T$;
  \item $P(\varphi(t,S_t,X_{t},Y_{t}) \ge 0 \forall t\le T)=1$
\end{enumerate}
where $\vf$ is the unique classical solution of \eqref{p1}-\eqref{18} in the prescribed class and $(\xi,\varepsilon)$ is as in \eqref{VI3.20}.

\noi In part (i) of Lemma \ref{lm2} it is shown that $\vf$ is a non-negative function. Hence (d) holds. From the definition of $\varepsilon_t$ in \eqref{VI3.20}, (c) follows. Next we show the condition (b). We apply It\^{o}'s formula to $e^{-\int_{0}^{t}r(X_{u})du} \varphi(t,S_{t},X_{t},Y_{t})$ under the
measure $P$ and use \eqref{8}, \eqref{p1} and \eqref{37}
to obtain after suitable rearrangement of terms
\begin{eqnarray*}\label{VI3.24}
e^{-\int_{0}^{t}r(X_{u})du} \varphi(t,S_{t},X_{t},Y_{t}) & = &
\varphi(0,S_{0},X_{0},Y_{0})+\int_{0}^{t}\frac{\partial
\varphi(u,S_{u},X_{u-},Y_{u-})}{\partial s} d S^*_{u}\no
+\int_{0}^{t}e^{-\int_{0}^{u}r(X_{v})dv}\\
& &\hspace{-5mm}\int_{\mathbb{R}} [\varphi(u,S_{u},X_{u-} +h(X_{u-},Y_{u-},z),Y_{u-}-g(X_{u-},Y_{u-},z))\\
&&-\varphi(u,S_{u},X_{u-},Y_{u-})]{\hat{\wp}}(du,dz)
\end{eqnarray*}
for all $t<T$. We set
\begin{eqnarray*}
L_t&:=& \int_{0}^{t}e^{-\int_{0}^{u}r(X_{v})dv}\int_{\mathbb{R}} [\varphi(u,S_{u},X_{u-} +h(X_{u-},Y_{u-},z),Y_{u-}-g(X_{u-},Y_{u-},z)) \\
&&-\varphi(u,S_{u},X_{u-},Y_{u-})]{\hat{\wp}}(du,dz).
\end{eqnarray*}
Since, $L_t$ is an integral w.r.t. a compensated Poisson random measure, it is a martingale. Again the independence of $W_t$ and $\wp$ implies the orthogonality of $L_t$ to the martingale part of $S^*_t$. Thus, we obtain the following F-S decomposition by letting $t \uparrow T $,
\begin{equation}\label{VI3.25}
B_T^{-1}(S_{T}-K)^+ = \varphi(0,S_{0},X_{0},Y_{0}) + \int^{T}_{0}{\xi_t}dS^*_t + L_T.
\end{equation}
Thus (a) and (b) hold.\qed
\begin{theo}\label{theo6}
Let $\varphi$ be the unique solution of \eqref{p1}-\eqref{18}. Set
\begin{eqnarray}\label{5}
\no\psi(t,s,i,y) &:=& \frac{ 1- F(T-t+y\mid i)}{1-F(y\mid i)}
\frac{\partial\eta_{i}(t,s)}{\partial s}+\int_0^{T-t}e^{-r(i)v}
\frac{f(y+v\mid i)} {1-F(y\mid i)} \times\sum_j p_{i j} \\
\no&& \int_0^{\infty} \varphi(t+v,x,j,0)
\frac{e^{\frac{-1}{2}((\ln(\frac{x}{s})-(r(i)
-\frac{\si^2(i)}{2})v) \frac{1}{\si(i)\sqrt v})^2}}{\sqrt{2\pi}
xs\sigma(i)\sqrt{v}}\frac{\left(\ln(\frac{x}{s})-(r(i)
-\frac{\si^2(i)}{2})v\right)}{\sigma(i)^2v} dx dv\\
\end{eqnarray}
where $(t,s,i,y)\in \mathcal{D}$.
Then $\psi(t,S_t,X_{t-},Y_{t-}) =\xi_t$ where $\xi_t$ is as in \eqref{VI3.20}.
\end{theo}

\proof We need to show that $\psi$ (as in \eqref{5}) is equal to $\frac{\partial\varphi}{\partial s}$. Indeed, one obtains the RHS of \eqref{5} by differentiating the right side of \eqref{34} with respect to $s$. Hence the proof. \qed

\begin{remark}\label{rem1}
 It is well known that in a numerical differentiation, an isolated perturbation gets enhanced whereas in a numerical integration that gets reduced. In \eqref{5}, the function $\psi$,
a partial derivative of $\varphi$, is given by a numerical integration involving $\varphi$. Thus the above theorem essentially provides a robust way to find the optimal hedging.
\end{remark}

\section{The risk measures associated to optimal hedging}
The \emph{quadratic residual risk} at $t=0$ associated with a strategy $\pi$, is denoted by $R_0(\pi)$ and is given by $R_0(\pi)=E[(C^*_T-C^*_0)^2|\mathcal{F}_0]$. In this section we compute this and some other closely related risk measures corresponding to the optimal strategy, discussed in the previous section. The following theorem is useful in this regard.
\begin{theo}\label{theo7}
Let $\{C_t\}$ be the accumulated additional cash flow process associated to the optimal hedging of the claim $H$ as in \eqref{4}. Then the quadratic variation process $[C]_t$ is given by
\begin{eqnarray*}
[C]_t = \sum_{r\in [0,t]} ( \varphi(r, S_r, X_r, Y_r)-\varphi(r, S_r, X_{r-}, Y_{r-}))^2
\end{eqnarray*}
where $\vf$ be the unique classical solution of \eqref{p1}-\eqref{18} with at most linear growth.
\end{theo}

\proof We have seen in Section 2 that the discounted value at $t=0$ of accumulated cash flow during $[0,T]$ is given by $$C^*_T = C^*_0+ \int_0^T \frac{1}{B_t} dC_t.$$ Further, from the F-S decomposition \eqref{VI3.25} we have obtained
\begin{eqnarray*}
L_T^{H^*} = \int_0^T \frac{1}{B_t} \int_{\mathbb{R}} ( \varphi(t, S_t, X_{t-}+h(X_{t-}, Y_{t-}, z),Y_{t-} - g(X_{t-}, Y_{t-},z))-\varphi(t, S_t, X_{t-}, Y_{t-})) \hat{\wp}(dt, dz).
\end{eqnarray*}
By comparing equations \eqref{1d} and \eqref{11a}, we get the relation $L_T^{H^*} = C^*_T - C^*_0$.
Hence we have
\begin{eqnarray}\label{eq: 3.3}
dC_t=\int_{\mathbb{R}}( \varphi(t, S_t, X_{t-}+h(X_{t-}, Y_{t-}, z),Y_{t-} - g(X_{t-}, Y_{t-},z))-\varphi(t, S_t, X_{t-}, Y_{t-}) ) \hat{\wp}(dt, dz).
\end{eqnarray}
Thus the external cash flow associated with the optimal hedging is obtained by integrating above. Hence
\begin{eqnarray*}\label{eq: 3.4}
C_T &=& C_0 + \int_0^T \int_{\mathbb{R}} ( \varphi(t, S_t, X_{t-}+h(X_{t-}, Y_{t-}, z), Y_{t-}-g(X_{t-}, Y_{t-}, z)) \nonumber\\
&&-\varphi(t, S_t, X_{t-},Y_{t-}))\hat{\wp}(dt, dz) \nonumber \\
&=& C_0 + \sum_{t\in [0,T]} ( \varphi(t, S_t, X_t, Y_t)-\varphi(t, S_t, X_{t-}, Y_{t-})) - \nonumber \int_0^T \sum_j \lambda_{X_{t-}j}(Y_{t-})
( \varphi(t, S_t, j, 0)\\
&&- \varphi(t, S_t, X_{t-}, Y_{t-}))dt.\quad \quad
\end{eqnarray*}
using $\hat{\wp}(dt,dz) =\wp(dt,dz) -dtdz$. Therefore, $C_t$ is an RCLL process.
For small $\Delta$ and $r\in(0,T)$
\begin{eqnarray*}
(C_{r}-C_{r-\Delta})^2&=&\Big(\varphi(r, S_r, X_r, Y_r)-\varphi(r, S_r, X_{r-\Delta}, Y_{r-\Delta})\Big)^2 -2 \Big( \varphi(r, S_r, X_r, Y_r) \nonumber\\
&&-\varphi(r, S_r, X_{r-\Delta}, Y_{r-\Delta})\Big) \sum_j \lambda_{X_{r-\Delta}j}(Y_{r-\Delta})\times \Big( \varphi(r, S_r, j, 0)\nonumber\\
&&-\varphi(r, S_r, X_{r-\Delta},Y_{r-\Delta})\Big)\Delta + \Big(\sum_j \lambda_{X_{r-\Delta}j}(Y_{r-\Delta})\big( \varphi(r, S_r, j, 0)\nonumber\\
&&-\varphi(r, S_r, X_{r-\Delta}, Y_{r-\Delta})\big)\Big)^2 \Delta^2.
\end{eqnarray*}
We recall that the quadratic variation process $[C]_t$ of $C_t$ is obtained by summing up the terms as in LHS over a partition of $[0,t]$ with $\Delta \to 0$. Hence the $\Delta^2$ terms as in the third term of RHS adds up to negligible. In the second term the coefficient of $\Delta$ converges to zero function except finitely many values of $r$ for almost every sample path. Hence the only significant term is the first one. Hence,
\begin{eqnarray*}
[C]_t = \sum_{r\in [0,t]} ( \varphi(r, S_r, X_r, Y_r)-\varphi(r, S_r, X_{r-}, Y_{r-}))^2.
\end{eqnarray*}\qed

\noi Although, from \eqref{eq: 3.3} one gets
\begin{eqnarray}\label{3.8}
C_T^* - C_0^* &=& \sum\limits_{t\in [0,T]} ( \varphi^*(t, S_t, X_t, Y_t)-\varphi^*(t, S_t, X_{t-}, Y_{t-})) - \int_0^T \sum\limits_j \lambda_{X_{t-}j}(Y_{t-}) ( \varphi^*(t, S_t, j, 0)\nonumber\\
&& - \varphi^*(t, S_t, X_{t-}, Y_{t-}))dt\nonumber \\
&=& \sum\limits_{n=1}^{n(T)} \Big\{ \varphi^*(T_n,S_{T_n},X_{T_n}, 0) - \varphi^*(T_n,S_{T_n},X_{T_{n-1}}, T_{n}-T_{n-1}) \nonumber \\
&&- \int_{T_{n-1}}^{T_n} \sum\limits_j \lambda_{X_{T_{n-1}}j}(t-T_{n-1})\lbrack \varphi^*(t,S_t,j,0)-\varphi^*(t,S_t,X_{T_{n-1}},t-T_{n-1})\rbrack dt \Big\} \nonumber \\
&&- \int_{T_{n(T)}}^T \sum\limits_j \lambda_{X_{T_n}j}(t-T_{n(T)}) \lbrack\varphi^*(t,S_t,j,0)-\varphi^*(t,S_t,X_{T_{n-1}},t-T_{n(T)})\rbrack dt
\end{eqnarray}
which can be used to deduce the expression of $R_0(\pi)$, but an application of It\^{o}'s isometry produces a simpler expression. Using It\^{o}'s isometry and Theorem \ref{theo7} we obtain
\begin{eqnarray}\label{qrr}
  R_0(\pi) &=& E [(C^*_T-C^*_0)^2|\mathcal{F}_0] \nonumber \\
  &=& E \bigg\lbrack\left(\int_0^T \frac{1}{B_t} dC_t\right)^2|\mathcal{F}_0\bigg\rbrack \nonumber \\
  &=& E\bigg\lbrack\int_0^T \frac{1}{{B_t}^2} d [C]_t|\mathcal{F}_0\bigg\rbrack \nonumber \\
  &=& E\bigg\lbrack\sum_{t\in [0,T]} \frac{1}{{B_t}^2}(\varphi(t, S_t, X_t, Y_t)-\varphi(t, S_t, X_{t-}, Y_{t-}))^2|\mathcal{F}_0 \bigg\rbrack \nonumber \\
  &=& E\bigg\lbrack\sum_{t\in [0,T]} ( \varphi^*(t, S_t, X_t, Y_t)-\varphi^*(t, S_t, X_{t-}, Y_{t-}))^2|\mathcal{F}_0 \bigg\rbrack \nonumber \\
  &=& E\bigg\lbrack\sum_{n=1}^{n(T)} \Big(\varphi^*(T_n, S_{T_n}, X_{T_n},0)-\varphi^*({T_n}, S_{T_n}, X_{T_{n-1}}, T_{n} - T_{n-1})\Big)^2|\mathcal{F}_0 \bigg\rbrack.
\end{eqnarray}
We would now compute some other closely related risk measures corresponding to the optimal strategy. To this end let us consider the following convex function $g:\mathbb{R}\rightarrow [0,\infty)$, given by $g(x)=x^2 1_{[0,\infty)}(x)$.
Let
\begin{eqnarray}\label{3.3a}
R_0^+(\pi):=E[g(C^*_T-C^*_0)|\mathcal{F}_0]
\end{eqnarray}
where, $C^*_t$ is the discounted cash flow associated with $\pi$. Clearly this is always less than the quadratic residual risk. This functional measures the amount of cash inflow unlike the QRR which gives a measure of cash flow in both the directions. We name this measure as positive residual risk (PRR). The value of $R^+_0$ for optimal hedging can be computed using \eqref{3.8}.

\noi It is also interesting to note that from \eqref{1d} one directly gets
\begin{eqnarray*}
C^*_T-C^*_0 = V_T^* - V_0 -\int_0^T \xi_t dS^*_t.
\end{eqnarray*}
In view of the fact that practitioners can trade assets only at discrete time intervals, here we assume that the writer trades at time $t_1<t_2< \dots<t_N$ and follows the optimal hedging suggestion obtained from the continuous time model. Thus the observed cash flow for this discrete trading is
\[C^*_T-C^*_0 = V_T^* - G_T^*\]
where \[G^*_T = V_0 +\sum\limits_{i=1}^N \xi_{t_i} \Delta S^*_{t_i}.\]
We call this variant as the \emph{practitioner's measure} (PM) of cash flow. This motivates us to introduce the following terms.
\begin{defn1}
The practitioner's measure of quadratic residual risk is given by
\begin{eqnarray}
PM(QRR)=E\lbrack (V^*_T - G^*_T)^2|\mathcal{F}_0 \rbrack.
\end{eqnarray}
and the practitioner's measure of positive residual risk is given by
\begin{eqnarray}
PM(PRR)=E\lbrack g(V^*_T - G^*_T)|\mathcal{F}_0 \rbrack.
\end{eqnarray}
\end{defn1}
\noi Computation of above measures are carried out in the next section by taking a typical market example.
\section{Numerical Method}
In this section we develop a robust method of computing the optimal hedging strategy using Theorem \ref{theo6}. There it is shown that the optimal strategy can be written in terms of two functions $\varphi$ and $\psi$, where $\psi$ is given by \eqref{5} provided $\varphi$ is known and $\varphi$ can be obtained by solving (\ref{34})-(\ref{35}).
We use a step-by-step quadrature method for finding a numerical approximation of the solution of integral equations (\ref{34})-(\ref{35}).
For a general study of quadrature method for linear integral equations we refer to \cite{BA}; \cite{BA1} and \cite{RA}. In view of \eqref{5} we need to compute $\varphi$ on $\{(t,s,i,y)\in \mathcal{D}|y=0\}$ only to compute $\psi$.

\noi Putting $y=0$ in (\ref{34}) we obtain
\begin{eqnarray}\label{ie0}
\no\varphi(t,s,i,0) &=&(1- F(T-t\mid i))
\eta_{i}(t,s)+\int_0^{T-t} e^{-r(i)v} f(v\mid i)\times\\
\no&& \sum_j p_{i j} \int_0^{\infty} \varphi(t+v,x,j,0)
\frac{e^{\frac{-1}{2}\left(\left(\ln(\frac{x}{s})-\left(r(i)
-\frac{\si^2(i)}{2}\right)v\right) \frac{1}{\si(i)\sqrt
v}\right)^2}} {\sqrt{2\pi} \si(i) x\sqrt v} dx dv.\\
\end{eqnarray}
Note that at $v=0$ the last integrand in (\ref{ie0}) is equal to $f(0\mid i)\sum p_{ij}\varphi(t,s,j,0)\Delta t.$
Therefore the dependence of the vector function $(\varphi(t, \cdot,
1,0),\varphi(t,\cdot,2,0),...,\varphi(t,\cdot,k,0))$ on its values at $t'\in (t,T]$ is
explicit. Thus even an implicit quadrature method to discretize
(\ref{ie0}) actually results in an explicit quadrature method
so we solve this in step-by-step manner with terminal condition
\begin{equation*}\label{iei0}
\varphi (T, s, i, 0)= (s-K)^+; \tab s\in \mathbb{R}; \tab i= 1, 2,
\ldots, k.
\end{equation*}

\noi Let $\Delta t$ be the time step and $\Delta s$ be the stock
step sizes respectively. For $m,m',l$ positive integers and $i
\in \mathcal{X}$ set
\begin{eqnarray*}
\mathcal{G}(m,m',l,i)&:=&
\frac{e^{\frac{-1}{2}((\ln(\frac{m'}{m})-(r(i)
-\frac{\si^2(i)}{2})l\Delta t) \frac{1}{\si(i)\sqrt{ l\Delta t}
})^2}}{\sqrt{2\pi}\si(i) m'\Delta s \sqrt{l\Delta t}},\\
\varphi_m^n(i) &\approx& \varphi(T- n\Delta t,m\Delta s,i,0).
\end{eqnarray*}
Now we use the following quadrature rule over successive
intervals $[0, n\Delta t]$: for a function $\psi$ on this
interval we use
$$\int_0^{n\Delta t} \psi(v) dv \approx \Delta t \sum_{l=0}^n
\omega_n(l)\psi(l)
$$
where $\omega_n(l)$ are weights to be chosen appropriately.
Applying the above discretization procedure in (\ref{ie0}) we
obtain the following set of equations
\begin{eqnarray}\label{q1}
\no \varphi_m^n(i) &=& (1-F(n\Delta t\mid i))\eta_{i}(T-n\Delta
t,m\Delta s) + \Delta t \sum_{l=1}^{n} \omega_n (l) e^{-r(i)
l\Delta t } f( l\Delta t\mid i)\\
&&\sum_j p_{i j}\Delta s \sum_{m'} \varphi_{m'}^{n-l}(j)
\mathcal{G}(m,m',l,i)+\Delta t \omega_n(0)f(0\mid i)\sum p_{ij}\varphi^n_m(j)
\end{eqnarray}
with
$$\varphi_m^0(i) = (m\Delta s-K)^+.
$$
We choose a repeated trapezium rule, that is, the weights
$\omega_n$ are given by
$$
\omega_n(l)= 1 \txt{ for } l=1,2,\ldots, n-1;~~
\omega_n(0)=\omega_n(n)=\frac{1}{2}.
$$
We now show that for sufficiently small $\Delta t$, the above
scheme (\ref{q1}) is \emph{strictly stable} with respect to an
isolated perturbation. We also show that the scheme displays
\emph{bounded error propagation} (i.e., the accumulated effect of
isolated perturbations $\delta$, added at each step, in
$\varphi_m^n(i)$, is uniformly bounded by a constant multiple of
$\delta$). We refer to \cite{BA1} for definitions and other
details.
\begin{theo}
Under (A2) let $a:= \displaystyle\max_{ \mathcal{X}\times [0,T]}
e^{-r(i)v} f(v\mid i)$. For
\begin{equation*}\label{dt}
\Delta t \le \frac{e^{-aT}}{a}
\end{equation*}
the scheme (\ref{q1}) is \emph{strictly stable} with respect to an
isolated perturbation. Moreover the scheme displays uniformly
bounded error propagation.
\end{theo}
\proof We first note that: (i) $\mathcal{G}(m,m',l,i)$
corresponds to a lognormal density, and (ii) under (A2) the
holding time densities $f(\cdot\mid \cdot)$ are bounded. Let
$\delta_n$ be an additive error in $\varphi_m^n(i)$ for all $m$
and $i$. Now it is easy to show that the effect of the isolated
perturbation $\delta_n$ in $\varphi_m^N(i)$ ($N := [\frac{T}
{\Delta t}]$) is
$$ \varepsilon_n = a \Delta t ( 1 + a\Delta t)^{N-n-1}\delta_n.
$$
If $\Delta t$ is sufficiently small satisfying (\ref{dt}) we get
$\varepsilon_n < \delta_n$, i.e., the scheme is strictly stable
with respect to an isolated perturbation. Let $\delta_n$ be
bounded by a fixed constant $\delta$. Now the total effects
$\varepsilon$ of the perturbations in the value $\varphi_m^N(i)$
is given by
$$ \varepsilon:=\sum_{n=0}^{N-1} \varepsilon_n ~
<~ (e^{aT}-1)\delta.
$$
Hence the result follows. \qed

\noi Having established the stability of the above scheme,
$\varphi_m^n(i)\approx $ $\varphi(T- n\Delta t,m\Delta s,i,0)$ is
computed for $n=0,1,2,\ldots, [\frac{T}{\Delta t}]; m=1,2,\ldots;
i=1,2,\ldots,k$, using the step-by-step quadrature method
(\ref{q1}). Next it is straightforward to compute $\varphi(T-
n\Delta t,m\Delta s,i,y)$ for a given $y$ using using the following discretization of \eqref{34}
\begin{eqnarray}\label{q2}
\no \varphi(T- n\Delta t,m\Delta s,i,y) &=& \frac{(1-F(n\Delta t
+ y \mid i))}{(1-F(y \mid i))}\eta_{i}(T-n\Delta t,m\Delta s)+ \Delta t \sum_{l=1}^{n} \omega_n (l) e^{-r(i)
l\Delta t } \\
 \no&& \frac{f( l\Delta t +y \mid i)}{(1-F(y \mid i))}\sum_j p_{i j}\Delta s \sum_{m'} \varphi_{m'}^{n-l}(j)
\mathcal{G}(m,m',l,i)\\
&&+\Delta t w_n(0)\frac{f(y|i)}{1-F(y|i)}\sum p_{ij}\varphi^n_m(j).
\end{eqnarray}
We now compute $\psi(T-
n\Delta t,m\Delta s,i,y)$ for a given $y$ using the following discretization of \eqref{5}
\begin{eqnarray}\label{q2.2}
\no \psi(T- n\Delta t,m\Delta s,i,y) &=& \frac{(1-F(n\Delta t
+ y \mid i))}{(1-F(y \mid i))}\frac{\partial\eta_{i}}{\partial s}(T-n\Delta t,m\Delta s)+ \Delta t \sum_{l=1}^{n} \omega_n (l) e^{-r(i)
l\Delta t } \\
 \no&& \frac{f( l\Delta t +y \mid i)}{(1-F(y \mid i))}\sum_j p_{i j} \sum_{m'} \varphi_{m'}^{n-l}(j)
\frac{\mathcal{G}(m,m',l,i)}{m}\left(\frac{\ln\left(\frac{m'}{m}\right)-(r(i)-\frac{\sigma^2(i)}{2})l\Delta t}{\sigma^2(i)l\Delta t}\right)\\
&&+\Delta t w_n(0)\frac{f(y|i)}{1-F(y|i)}\sum p_{ij}\frac{\varphi^n_m(j)}{m\Delta s}\left(\frac{1}{2}-\frac{r(i)}{\sigma^2(i)}\right).
\end{eqnarray}
The system of equations \eqref{q2.2} gives a numerical approximation for
the optimal hedging strategy corresponding to the European call option.

\noi For illustration of the results we next consider an example of a
semi-Markov modulated market with three regimes. The state space
$\mathcal{X}$ is $\{1,2,3\}$. The drift coefficient, volatility and
instantaneous interest rate at each regime are chosen as follows
$$\Big(\mu(i), \si(i), r(i)\Big):=\left\{
\begin{array}{ccc}
  (0.2,0.2,0.2) & \txt{if} & i=1~~~ \\
  (0.6,0.4,0.5) & \txt{if} & i=2~~~\\
  (0.8,0.3,0.7) & \txt{if} & i=3~~.
\end{array}\right.$$
The transition probability matrix is assumed to be given by
$$(p_{ij})= \left(\begin{array}{ccc}
 0 & 2/3 & 1/3 \\
  1/2 & 0 & 1/2 \\
  1/3 & 2/3 & 0
\end{array}\right).$$
In this example the holding time in each regime is assumed to be
$\Gamma(2,1)$. That is
$$f(y\mid i) = ye^{-y},~ y\ge 0 \txt{~and~} i =1,2,3.$$

\begin{figure}[h]
  \centering
 \includegraphics[width=0.5\textwidth]{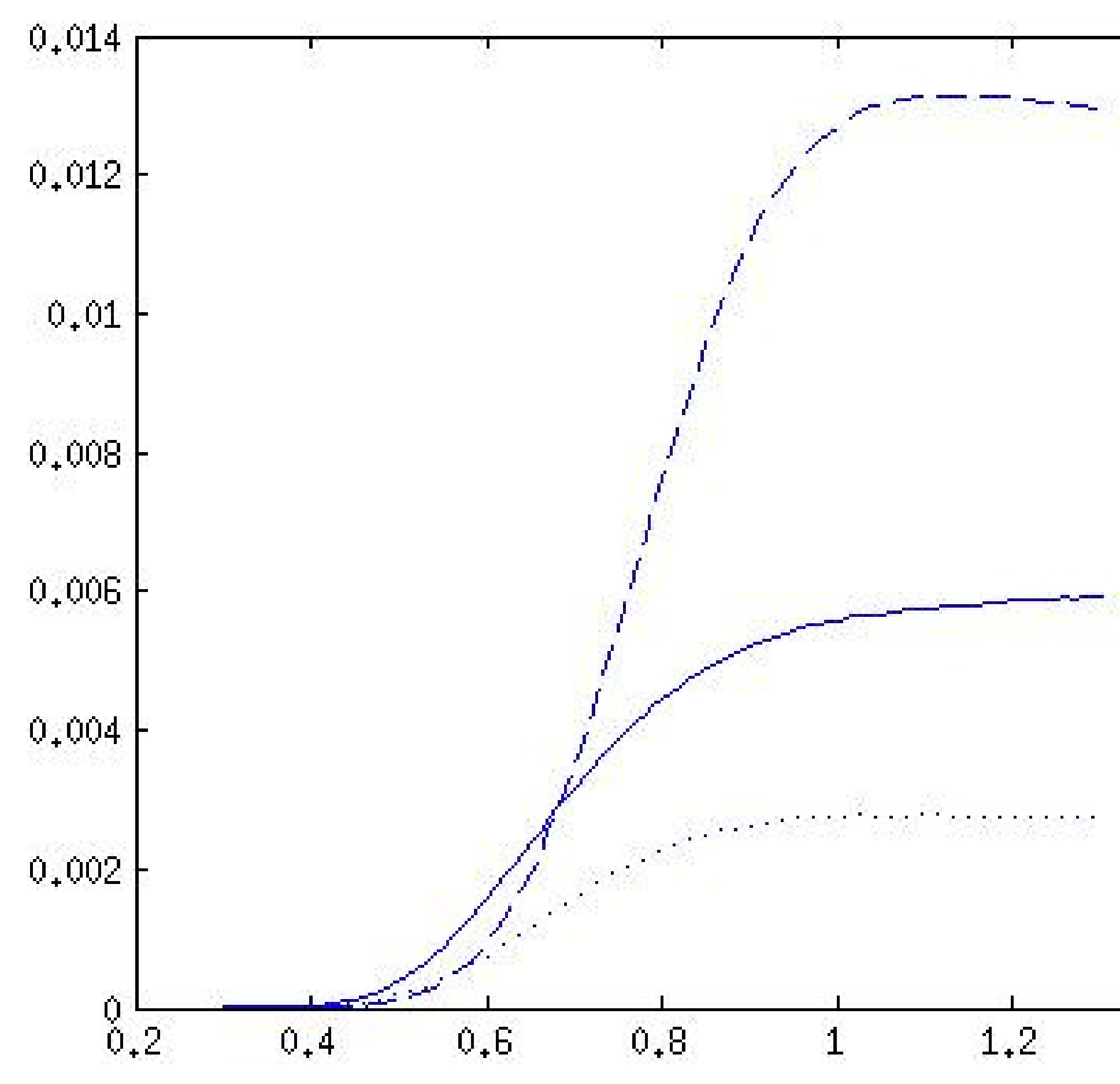}
  \caption{QRR}\label{fig1}
\end{figure}
In this particular market model we compute the QRR associated to the optimal hedging using \eqref{qrr} for a European call option with $T=1, K=1$. This requires prior knowledge of the option price function. We compute that numerically using \eqref{q2} for this semi-Markov modulated market.
The expression in \eqref{3.8} also involves a conditional expectation. We have taken 101 equi-spaced
grid points on the interval $[0.3,1.3]$ which also includes the strike
price $K=1$. For each grid point $s$ and each $i= 1,2,3$ we
compute the conditional expectation by simulating the process
$(S_t, X_t, Y_t)$ $10^6$ times starting with $S_0=s, X_0=i, Y_0=0$ with time step size, $\Delta t=$. In the Figure \ref{fig1} for each $i$
we plot the function $R_0(\pi)(s,i)$ along the vertical axis against $S_0=s$ along the
horizontal axis. The plots of $R_0(\pi)(s,1)$, $R_0(\pi)(s,2)$
and $R_0(\pi)(s,3)$ are put together in one frame for clear
comparison. One obvious observation is that due to incompleteness
of the market the quadratic residual risk at $t=0$ is nonzero. 
Beside this, the plot leads to another important observation
regarding relative behavior of $R_0(\pi)$ at different regimes.
\begin{figure}[h]
  \centering
 \includegraphics[width=0.5\textwidth]{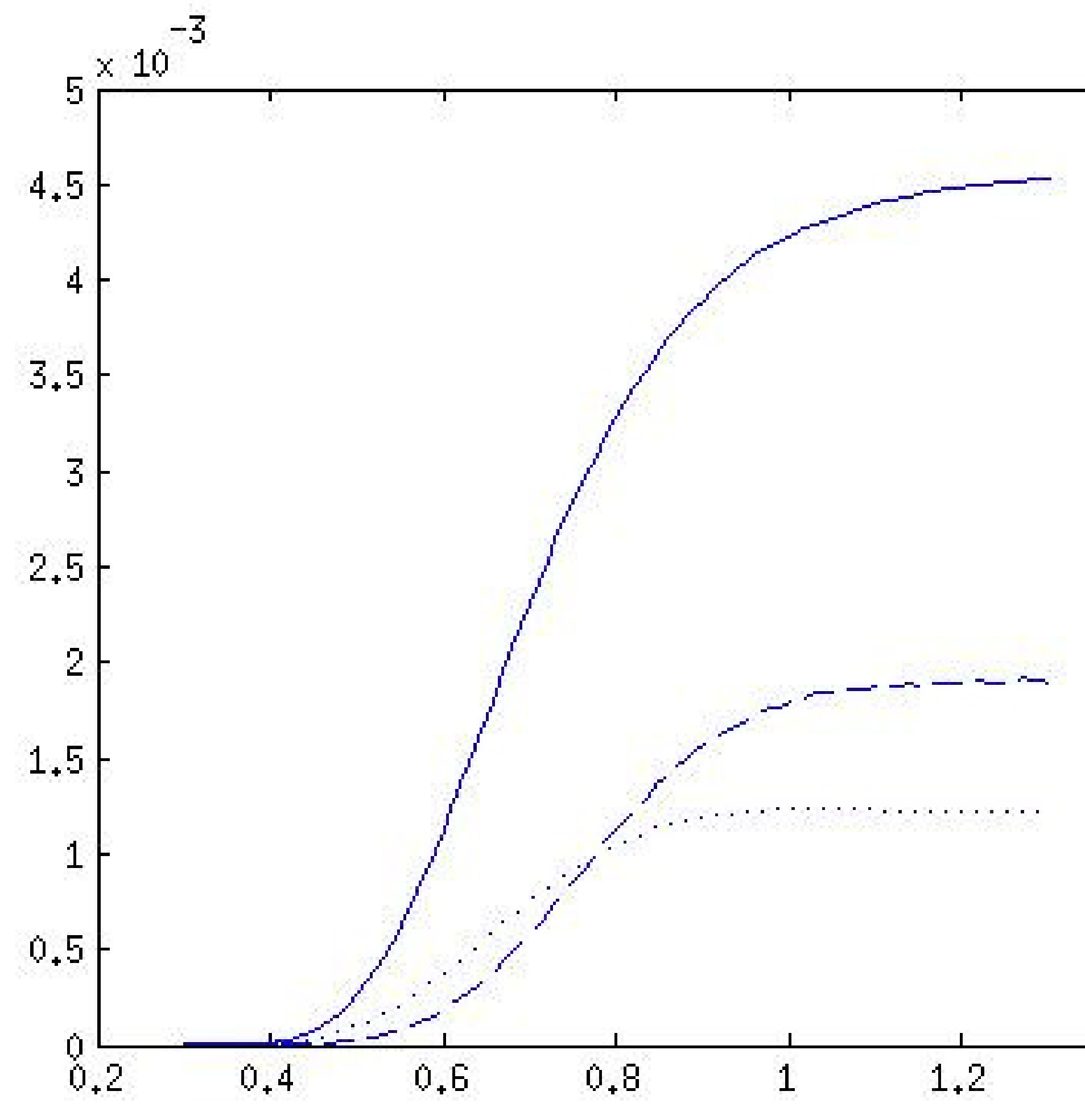}
  \caption{PRR}\label{fig2}
\end{figure}

Next we compute PRR as in \eqref{3.3a} for above example using \eqref{3.8}. The Figure \ref{fig2} shows PRR values
for three different regimes and various different initial stock prices ranging from 0.3 to 1.3.
A clear comparison between Figure \ref{fig1} and Figure \ref{fig2} exhibits the fact that PRR is alway less than QRR.

Unlike QRR and PRR, the measures PM(QRR) and PM(PRR) do not have explicit dependence on option price but these depend on the hedging. In Theorem \ref{theo6} we have obtained an expression of hedging in terms of price function. Using that expression one can compute the optimal hedging robustness of this method is discussed in Remark \ref{rem1}. Now we compute PM(QRR) and PM(PRR) respectively for the above mentioned market example using \eqref{5} where the option prices are obtained using the numerical scheme \eqref{q2}. Figure 3 and Figure 4 show the respective values of PM(QRR) and PM(PRR) at different initial regimes and different initial stock prices.
\begin{figure}[h]
  \centering
 \includegraphics[width=0.5\textwidth]{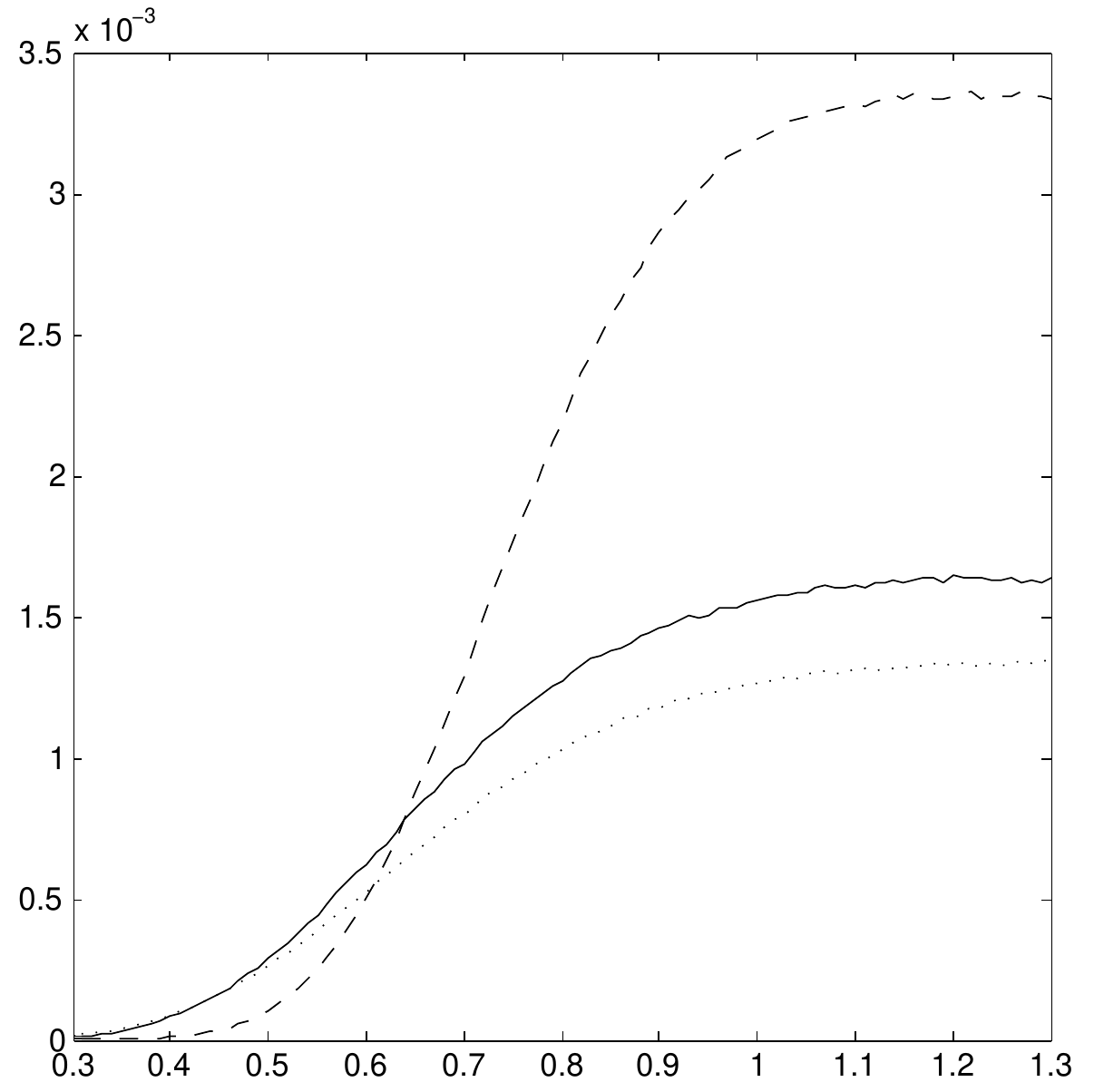}
  \caption{PM(QRR)}\label{fig3}
\end{figure}\begin{figure}[h]
  \centering
 \includegraphics[width=0.5\textwidth]{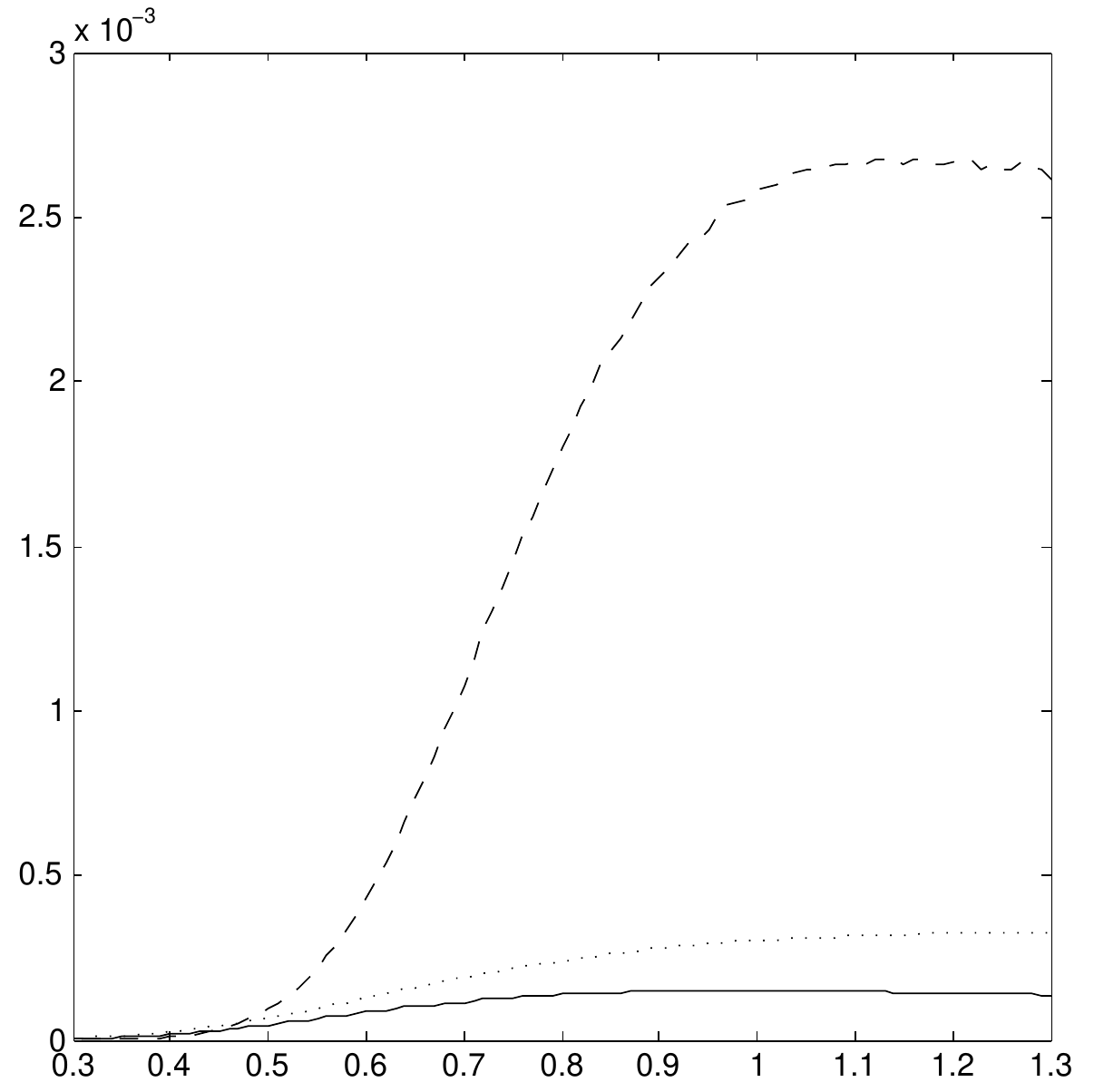}
  \caption{PM(PRR)}\label{fig4}
\end{figure}

\end{document}